\documentclass[prb,aps,twocolumn]{revtex4}

\usepackage[dvips]{graphics}
\usepackage{psfig}
\usepackage{graphicx}
\usepackage{amssymb}
\usepackage{epstopdf}

\begin{document}

\title{Ferromagnet-superconductor proximity effect: The clean limit}
\author{Milo\v{s} Bo\v{z}ovi\'{c} and Zoran Radovi\'c}
\address{Department of Physics, University of Belgrade, P.O. Box 368, 11001 Belgrade, Serbia and Montenegro}


\begin{abstract}
We study theoretically the influence of ferromagnetic
metals on a superconducting film in the clean limit. Using a
self-consistent solution of the Bogoliubov--de Gennes equation for
a ferromagnet-superconductor-ferromagnet double junction we
calculate the pair potential and conductance spectra
as a function of the superconducting layer thickness $d$ for
different strengths of ferromagnets and interface transparencies.
We find that the pair potential and the critical temperature are
weakly perturbed by the exchange interaction and do not drop
to zero for any finite $d$. On the other hand, for thin superconducting films
charge transport is spin polarized and
exhibits a significant dependence on the ferromagnetic strength
and magnetization alignment.
\end{abstract}

\maketitle

\section{Introduction}

Rapid advancement of nanofabrication technology in the past decade
has reinvigorated interest in understanding the effects inherent
to clean superconducting heterostructures. Apart from potential
device applications, particularly in quantum information storage
and processing,\cite{Oh} variety of phenomena makes
systems consisting of superconductor sandwiched between two ferromagnets
interesting for both experimental and theoretical investigation
(see, for example,
refs.~\cite{Lazar,Aarts,GuPratt,Gu,aa,Garifullin,Obi} and
~\cite{beasley,Valls01,Buzdin01,Fominov,BozovicB,Dong,Yamashita,Buzdin03,Bagrets,You,Valls04,Melin}).
The interplay of ferromagnetism and superconductivity results in
characteristic proximity effects near the contacts between two
metals: the Andreev reflection is suppressed in the presence of a
ferromagnet;\cite{dJB} conversely, the superconducting
correlations, described via the pair amplitude, extend in the
ferromagnetic material in an oscillatory
way.\cite{Valls01}

The proximity effect in heterostructures consisting of a
superconductor (S) in contact with a normal, nonmagnetic (N) or
ferromagnetic (F), metal has been widely studied and well
understood in the dirty limit, where the electron mean free path
$l$ is much smaller than the superconducting coherence length
$\xi_0$.\cite{Deut,Radovic91,jiang}
In particular, a strong depairing effect is found in FSF
trilayers, which results in a superconducting-to-normal phase
transition at a finite thickness of the S layer.\cite{Aarts,Lazar}
Furthermore, using a spin-valve setup for these junctions,
spontaneous transition from parallel (P) to antiparallel (AP)
ferromagnetic moment orientation could be
triggered.\cite{Buzdin03}

However, in recent experiments with FSF trilayers it was found that the difference
between the superconducting critical temperatures for AP and P alignment is less by two
orders of magnitude than theoretical predictions for the dirty
limit.\cite{Gu,aa} A possible reason for this disagreement could
be that these samples were actually much cleaner than the ones
used in previous experiments.\cite{Lazar,Aarts}
Namely, as a consequence of
phase coherence of electron wave functions
quasiparticle spectrum of clean trilayers
differs substantially from the BCS result for bulk superconductor:
density of states is practically gapless and Andreev reflection is reduced in thin S layers.\cite{BozovicB,Yamashita}
As a result, depairing induced by ferromagnets appears to be much weaker in the clean
limit, when $l\gg\xi_0$.\cite{Valls01}

The purpose of this paper is to clarify the
ferromagnet-superconductor inverse proximity effect in the clean limit,
within the framework of the BCS theory. The model we study is an FISIF
heterostructure with an S film in contact with massive F
metals, where I denotes the interface potential barrier of
arbitrary transparency. To describe the equilibrium and transport
properties of such a system, we use a numerical procedure to solve
the Bogoliubov--de Gennes equation self-consistently for
spatial-averaged pair potential. We find that neither the
magnitude of exchange interaction in the ferromagnets nor the
relative orientation of the magnetizations has any significant
influence on the pair potential and
the critical temperature of the superconductor.
On the other hand, transport properties of thin S films exhibit a significant
dependence on ferromagnetic strength and alignment of
magnetizations.\cite{BozovicB,Yamashita} Unlike the
case of diffusive mesoscopic superconducting bilayers and
trilayers,\cite{FF} we found a BCS type of the pair potential
temperature dependence. A peculiar temperature and alignment
dependence of the proximity effect is recently obtained for a
model of FSF trilayers with atomic
thickness.\cite{Daumens,Feinberg}

\section{Equilibrium properties}

\begin{figure}[h]
\includegraphics[width=5cm]{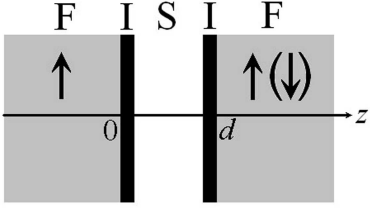}
    \caption{The geometry of
    FISIF heterostructure in P (AP) alignment of magnetizations.}
    \label{f.1}
\end{figure}

In this section we study the equilibrium properties of an $s$-wave
superconductor sandwiched in between two ferromagnets. We consider a simple model of FISIF double junction consisting of
a superconducting layer of thickness $d$ connected to
ferromagnetic metals by interfaces of arbitrary transparency, fig.~\ref{f.1}.
Assuming that the metals are clean, quasiparticle
propagation is described by the Bogoliubov--de Gennes equation
\begin{eqnarray}
\label{BdG}
\left(
\begin{array}{cc}
  H_0({\bf r})-\rho_{\sigma}h({\bf r}) & \Delta({\bf r}) \\
  \Delta^\ast({\bf r}) & -H_0^\ast({\bf r})+\rho_{\bar{\sigma}}h({\bf r})
\end{array}
\right) \left(\begin{array}{c}
    u_\sigma({\bf r}) \\
    v_{\bar{\sigma}}({\bf r})
  \end{array}\right) = \nonumber\\ = E\left(\begin{array}{c}
    u_\sigma({\bf r}) \\
    v_{\bar{\sigma}}({\bf r})
  \end{array}\right),
\end{eqnarray}
where $H_{0}({\bf r})=-\hbar^{2}\nabla^{2}/2m+W({\bf r})-E_F$,
$E_F$ is the Fermi energy, $\sigma$ is the quasiparticle spin
orientation ($\sigma =\uparrow, \downarrow$ and
$\bar{\sigma}=\downarrow, \uparrow$), and $E$ is the quasiparticle
energy with respect to the Fermi level. The interface potential is
modeled by $W({\bf r})=\hat{W}\{\delta(z)+\delta(z-d)\}$, where
the $z$ axis is perpendicular to the layers and $\delta(z)$ is the
Dirac delta-function. For simplicity, the electron effective
mass $m$ and the Fermi velocity, $v_F=\sqrt{2mE_F/\hbar^2}$, are
assumed to be constant through the junction. The superconducting
pair potential is taken in the form $\Delta({\bf r})=\Delta(z)
\Theta(z) \Theta(d-z)$, where $\Theta(z)$ is the Heaviside step
function. The exchange potential $h({\bf r})$ is given by
$h_{0}\{\Theta(-z)+[-]\Theta(z-d)\}$ for the P [AP] alignment, and
$\rho_{\sigma}$ is $1$ ($-1$) for $\sigma=\uparrow$
($\downarrow$); a uniform magnetization is assumed to be parallel to the
layers. The parallel component of the wave vector, ${\bf
q}_{||}\equiv\left(q_x, q_y, 0\right)$, is conserved, and the
spinor $\left( u_\sigma({\bf r})~~v_{\bar{\sigma}}({\bf r})
\right)^{\rm T}$ satisfies the appropriate boundary conditions for
the wave function and its first derivative at $z=0$ and
$z=d$.\cite{BozovicB}

For a constant pair potential, solution of eq.~(\ref{BdG})
has the following form in the superconductor, $0<z<d$,
\begin{eqnarray}
\label{sol}
\left(\begin{array}{c}
    u_\sigma({\bf r}) \\
    v_{\bar{\sigma}}({\bf r})
  \end{array}\right) = \exp(i{\bf q_\parallel}\cdot{\bf r}) \times
  ~~~~~~~~~~~~~~~~~~~~~~~~~~~~~~~~~~~~~~\\
\times\Big\lbrace
\left[c_{1}(E,{\bf q_\parallel})\exp(iq^+_z z)
+c_{2}(E,{\bf
q_\parallel})\exp(-iq^+_z z)\right] \left(\begin{array}{c}
    \bar{u} \\
    \bar{v}
  \end{array}\right) \nonumber \\
+ \left[ c_{3}(E,{\bf q_\parallel})\exp(iq^-_z z)+c_{4}(E,{\bf
q_\parallel})\exp(-iq^-_z z)\right]\left(\begin{array}{c}
    \bar{v} \\
    \bar{u}
  \end{array}\right)\Big\rbrace.\nonumber
\end{eqnarray}
Here, $\bar{u}=\sqrt{(1+\Omega/E)/2}$ and
$\bar{v}=\sqrt{(1-\Omega/E)/2}$ are the BCS coherence factors,
$\Omega=\sqrt{E^2-\Delta^2}$, and the modulus of the wave vector
${\bf q}^\pm \equiv \left(q_x, q_y,q_z^\pm \right)$ is given by
\linebreak $|{\bf q}^\pm|=\sqrt{(2m/\hbar ^2)(E_F\pm\Omega)}$.
Coefficients $c_1$ through $c_4$ are obtained from boundary
conditions in the scattering problem for FISIF
heterostructure.\cite{BozovicB}

Wave-vector components have to be normalized by using the
condition for canonical transformation
\begin{eqnarray}
\label{norm} \int_V {\rm d}^3{\bf r} \left[ |u_\sigma ({\bf r})|^2+
|v_{\bar{\sigma}} ({\bf r})|^2 \right] = 1,
\end{eqnarray}
where the integration is performed over the volume $V$ of the superconductor. The
self-consistency condition for the pair potential is given by\cite{Ket}
\begin{equation}
\label{self2} \Delta(z)=\lambda \sum_{\bf q}\left\{ u_\uparrow
({\bf r}) v^\ast_\downarrow ({\bf r})\left[1-f_0(E)\right] -
u_\downarrow ({\bf r}) v^\ast_\uparrow ({\bf r})f_0(E)
\right\}.
\end{equation}
Here, $\lambda$ is the coupling constant and $f_0$ is the Fermi
distribution function at temperature $T$. By
performing the summation over ${\bf q}$, we get
\begin{eqnarray}
\label{self}
&~&\Delta(z)=\lambda N(0) V \int_0^{\pi/2} {\rm
d}\theta
\sin\theta\cos\theta \times\nonumber\\ &\times&\int_0^{\hbar\omega_D}{\rm d}\Omega
\left\{ u_\uparrow ({\bf r}) v^\ast_\downarrow ({\bf
r})\left[1-f_0(E)\right] - u_\downarrow ({\bf r})
v^\ast_\uparrow ({\bf r})f_0(E) \right\},\nonumber
\end{eqnarray}
where $N(0)=mk_F/2\pi^2\hbar^2$ is the normal-metal density of
states (per spin orientation) at the Fermi level, $\theta$ is the
angle between ${\bf q}^+$ and the $z$ axis, and $\hbar\omega_D$ is
the upper cutoff in integration over quasiparticle kinetic energy
$\Omega$.
\begin{figure}[htb]
\includegraphics[width=7.5cm]{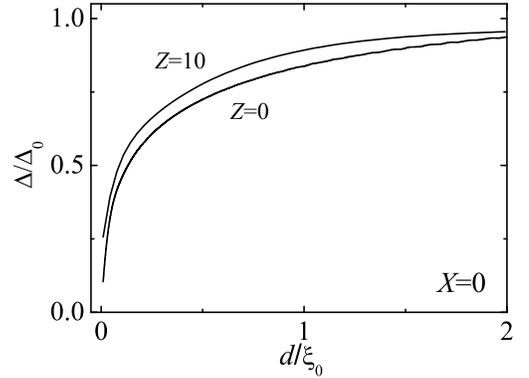}
\caption{ Zero-temperature pair potential $\bar{\Delta}$,
            normalized to the bulk value $\Delta_0$,
            as a function of the S film thickness $d$ for NSN ($Z=0$) and NISIN ($Z=10$)
            junctions. The same curves describe dependence
            of $T_c/T_{c0}$ on $d$.}
\label{f.2}

\end{figure}

\begin{figure}[htb]
\includegraphics[width=7.5cm]{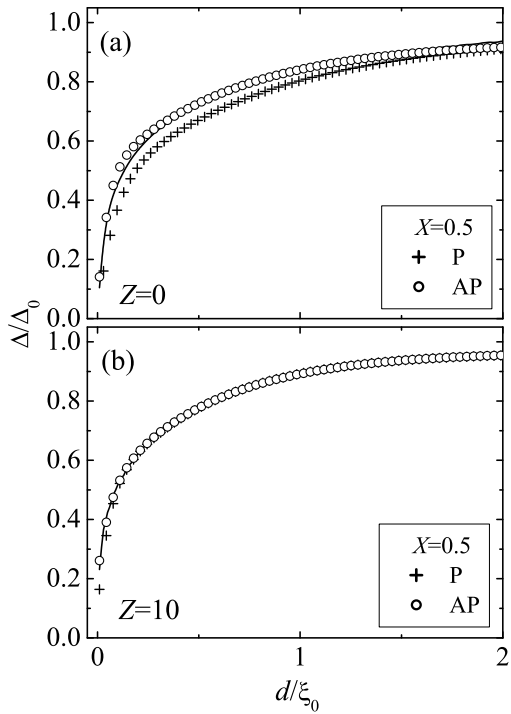}
\caption{ Zero-temperature pair potential $\bar{\Delta}$,
normalized to the
            bulk value $\Delta_0$, for (a) FSF ($Z=0$) and (b) FISIF ($Z=10$)
            junctions as a function of the S film thickness $d$ for $X=0.5$ and
            for parallel and antiparallel magnetization alignments. The corresponding
            NISIN curves ($X=0$) are shown for comparison (solid lines).
            The same curves describe dependence of $T_c/T_{c0}$ on $d$.}
\label{f.3}
\end{figure}

We solve eq.~(\ref{BdG}) using the stepwise approximation for $\Delta(z)$.
We calculate the spatial average of $\Delta(z)$ following a
standard iteration procedure
\begin{equation}
\label{aver} \bar{\Delta}_{i+1}=
\frac{1}{d}\int_0^d\Delta_i(z){\rm d}z
\end{equation}
by setting $E=\sqrt{\Omega^2+\bar{\Delta}^2_i}$ in eq.~(\ref{self2}) in order to obtain $\Delta_i(z)$ in the $i$-th
iteration.\cite{slatt} Starting from the bulk value
$\Delta_0$ we repeat this procedure until the difference between
$\bar{\Delta}_{i+1}$ and $\bar{\Delta}_i$ becomes sufficiently
small. This procedure is justified both for thin S films, $d/\xi_0\lesssim 1$,
where the spatial variation of the pair potential is small,
and thick S films, $d/\xi_0\gg 1$, where $\Delta(z)$ is practically flat at $\Delta_0$,
except in a narrow region of the order of $\xi_0$ near the interfaces.

Numerical results obtained by the self-consistency algorithm
described above are shown in figs.~\ref{f.2} and \ref{f.3}. The
average pair potential $\bar{\Delta}$, normalized to the bulk
value $\Delta_0$, is shown in fig.~\ref{f.2} as a function of
$d/\xi_0$, where $\xi_0=\hbar v_F/\pi\Delta_0$, for an NISIN
junction at zero temperature and for two strengths of the
interface barriers, $Z\equiv 2\hat{W}/\hbar v_F$. As the transmissivity
of the interfaces decreases ({\it i.e.}, as $Z$ increases),
the normal reflection of incoming electrons becomes more probable, which makes the S film more isolated and weakens the proximity
effect. This could be seen for $Z=10$, which corresponds to
a single-barrier transmissivity of $\sim1\%$ of the electrons
injected into the superconductor. In the case of
transparent interfaces, $Z=0$, the pair potential is more
suppressed by the proximity of the normal-metal electrodes.

We have studied ferromagnetic influence for various strengths of
the normalized exchange potential $X\equiv h_0/E_F$ in the F
metals for both P and AP magnetization alignment.\footnote{The case of arbitrary
orientation of ferromagnetic moments requires the introduction of triplet components
of the anomalous Green's function.\cite{Bergeret}} In fig.~\ref{f.3} we
show the particular case of $X=0.5$, for FSF ($Z=0$) and FISIF
($Z=10$) heterostructures. It can be seen that all the points,
calculated by numerical integration, fall very close to the $X=0$
curves, almost independently of the magnetization alignment. This
is in a strong contrast with dirty FSF structures, which exhibit a
phase transition to normal state at some critical value of the
superconducting film thickness\cite{Aarts,Lazar} which
varies with relative orientation of the
magnetizations.\cite{Buzdin01,Buzdin03} We conclude that in clean
FISIF systems the superconducting pair potential is practically
independent of ferromagnetic polarization. This
result is in accord with the previous ones obtained by a fully
self-consistent description of the proximity effect at the F-S
interface.\cite{Valls01} Recent experiments also imply that the
difference between the superconducting critical temperatures in AP and
P alignment is off by about two orders of magnitude with respect to the theoretical
predictions for the dirty limit.\cite{Gu,aa} A possible reason for
this disagreement could be that these samples were much cleaner
than what was the case in earlier experiments (refs.~\cite{Lazar}
and \cite{Aarts}) where data fit very well to the dirty-limit
theory.

Qualitatively, the fact that $\bar{\Delta}$ weakly depends on $X$
in the clean limit can easily be shown using the quasiclassical
approximation.\cite{Andreev,Sanchez} By neglecting the wave vectors outside the small
interval around the Fermi surface of radius $k_F$, so that
$\hbar^2 {\bf q}^2/2m \simeq -E_F\pm \hbar^2 k_F |{\bf q}|/m$, the
one-electron Hamiltonian could be linearized, $H_0 \simeq \pm i
(\hbar^2 k_F/m) (\partial/\partial z)-2E_F$. After normalization given by eq.~(\ref{norm}),
coefficients in eq.~(\ref{sol}) become
\begin{eqnarray*}
\label{c1c4norm}
c_1(E,{\bf q_\parallel})=\frac{\bar{u}}{\sqrt{V\left(1-\bar{u}^2\bar{v}^2 \zeta^{-1}\sin\zeta \right)}}, &~~~&
c_2(E,{\bf q_\parallel})=0,\nonumber\\
c_3(E,{\bf q_\parallel})=-\frac{\bar{v}
e^{i\zeta}}{\sqrt{V\left(1-\bar{u}^2\bar{v}^2 \zeta^{-1}\sin\zeta
\right)}},&~~~& c_4(E,{\bf q_\parallel})=0,
\end{eqnarray*}
where $\zeta \equiv d\left(q^+_z -
q^-_z\right)$. Therefore, solutions $u_\sigma({\bf r})$ and $v_\sigma({\bf r})$
of linearized eq.~(\ref{BdG}) are independent of $X$ and exactly match the
solutions of the Bogoliubov--de Gennes equation for transparent
NSN junctions. We emphasize that a weak influence of the exchange potential
is inherent for clean S and F metals. The dependence of the critical temperatures on the mutual orientations of ferromagnetic moments is also hardly observable in dirty FISIF hybrids with finite interface transparency.\cite{Buzdin03,You} However, for such structures with higher transparency the inverse proximity
effect is significant.\cite{Lazar,Aarts}

We have also verified that the temperature dependence of
$\bar{\Delta}$ remains to be of the BCS type,\cite{muhl} well
described by $\bar{\Delta}(T)=\bar{\Delta}(0)
\tanh\left(1.74\sqrt{T_c/T-1}\right)$ for all $d$, where $T_c$ is
the critical temperature of the S film. In addition, the
$d$-dependence of the critical temperature normalized to the bulk
value, $T_c/T_{c0}$, coincides with $\bar{\Delta}(0)/\Delta_0(0)$
vs. $d$ curves (figs.~\ref{f.2} and \ref{f.3}), being greater for AP than for P alignment of magnetizations, which is in agreement with theoretical predictions\cite{You,Buzdin03} and experimental results\cite{Gu}. In diffusive
mesoscopic superconducting bilayers and trilayers, however, a more
complex dependence is predicted.\cite{FF} Moreover, for finite F layers
the critical temperature, as well as the pair potential, oscillate with the F-layer
thickness.\cite{Gu,Garifullin,Bagrets,Obi}


\section{Transport properties}

\begin{figure}[h]
\includegraphics[width=7.cm]{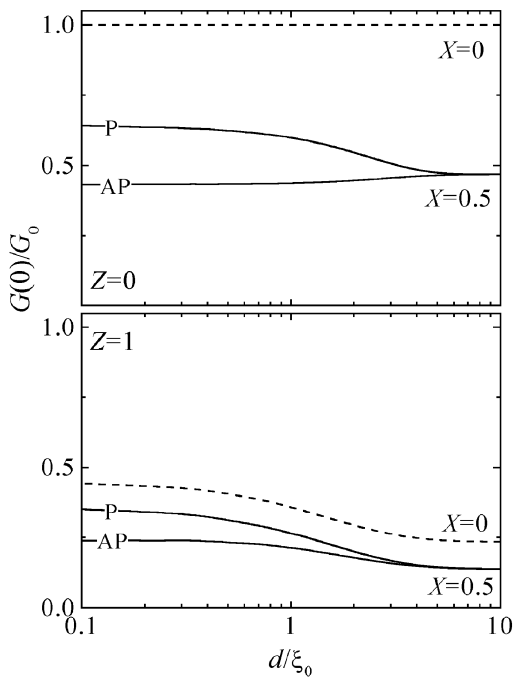}
\caption{ Zero bias conductances of FSF ($Z=0$, top panel) and
FISIF
            ($Z=1$, bottom panel) junctions at zero temperature as a function of the S film
            thickness $d$ for $X=0.5$. The corresponding NSN and NISIN spectra ($X=0$, dashed
            lines) are shown for comparison.}
\label{f.4}
\end{figure}

The properties of quasiparticle transport are commonly
described by differential conductance spectrum. For an FISIF
double-barrier junction at zero temperature conductance per orbital transverse channel
can be calculated from\cite{Lambert,Dong,Yamashita}
\begin{eqnarray}
\label{conduct} G(E) &=& G_0\sum_{\sigma=\uparrow,\downarrow}
P_\sigma \times\nonumber\\
&\times&\int_0^{\pi/2} {\rm d}\theta\sin\theta\cos\theta
~\left[A_\sigma(E,\theta)+C_\sigma(E,\theta)\right],
\end{eqnarray}
where $G_0=2e^2/h$ is the conductance quantum and $P_\sigma=\left(1+\rho_\sigma X\right)/2$.
Analytical results for the probabilities of Andreev reflection and
direct quasiparticle transmission, $A_\sigma(E,\theta)$ and
$C_\sigma(E,\theta)$, respectively, are presented and
discussed in ref.~\cite{BozovicB}. Contribution of evanescent
propagation to $G(E)$ is included as well. Here, we only point out
the strong influence of ferromagnetism on charge transport through
a thin S film, in contrast to the weak proximity effect on
equilibrium properties.

As an illustration, we calculate the zero bias conductance,
$G(0)$, using the self-consistent pair potential obtained in the
previous section. The results are shown in fig.~\ref{f.4}.
The subgap transport through the superconductor
changes by virtue of two principal mechanisms: firstly, due to the
presence of the ferromagnets, and secondly, due to the decrease of
thickness $d$. The zero-bias conductance
for an FSF junction with $X=0.5$ is significantly
below the NSN value of unit conductance per channel,
and splits for the P and the AP magnetization
alignment as the S film becomes thinner. The subgap conductance is greater in
P than in the AP alignment as a result of a strong magnetoresistive effect in
thin S layers: when $d\lesssim\xi_0$ the direct transmission of spin polarized quasiparticles
across the superconductor becomes a dominant transport mechanism. \cite{BozovicB,Yamashita}
Hence, in contrast to the NSN trilayers, the conductance of an FSF system
shows a strong size
effect. Conductances become significantly suppressed for finite
interface transparency ($Z=1$), both for NISIN and FISIF
junctions, due to the increase of normal and decrease of
Andreev reflection probability.


\section{Conclusion}

We have studied the ferromagnet-superconductor proximity effect in
clean FISIF heterostructures with thin S layers and massive F
metals. We have found that the superconducting order parameter is
weakly affected by the exchange interaction and has practically
the same dependence on the S film thickness as in the
corresponding NISIN structures.
On the other hand, quasiparticle dynamics within an FISIF heterojunction
is substantially changed with respect to the corresponding NISIN
system, the more so as the S film becomes thinner.

While in clean ferromagnet-superconductor
hybrids with massive F layers the ferromagnetism
has negligible inverse proximity effect on equilibrium properties, the
opposite behavior is previously found for the dirty ones,\cite{Aarts}
where the pairing potential may be very sensitive
to the vicinity of the F layer(s). Moreover, in dirty hybrids the critical value of
superconducting layer thickness at which destruction of
superconductivity occurs is strongly dependent on the
ferromagnetic exchange potential.\cite{Buzdin03} On the other hand, charge
transport is diffusive in this case, and consequently the conductance
spectrum is practically independent on the S film thickness and alignment of
magnetizations.

In summary, we have found that in the clean limit the BCS
self-consistent solution for a thin superconducting film in
contact with massive ferromagnets shows very weak
depairing effect of the exchange interaction. While the
equilibrium properties are practically unaffected, charge
transport is strongly influenced by proximity of ferromagnets.

\bigskip
\section*{ACKNOWLEDGMENT}

We are grateful to N. Chtchelkachev, L. Tagirov, and I. Bo\v zovi\'c for useful
discussions. The work has been supported by the Serbian Ministry
of Science, Project No. 1899.

\end{document}